\documentstyle[12pt]{article}

\newcommand{\EQ}{\begin{equation}}
\newcommand{\EN}{\end{equation}}
\def\aprle{\buildrel < \over {_{\sim}}}
\def\aprge{\buildrel > \over {_{\sim}}}
\begin{document}
\topmargin 0pt
\oddsidemargin=-0.4truecm
\evensidemargin=-0.4truecm
\renewcommand{\thefootnote}{\fnsymbol{footnote}}
\newpage
\setcounter{page}{0}
\begin{titlepage}
\vspace{0.8cm}
\begin{flushright}
SISSA -172/92/EP, ~Roma n. 902\\
\end{flushright}
\vspace{0.3cm}
\begin{center}
{\large IMPLICATIONS OF GALLIUM SOLAR NEUTRINO DATA FOR THE
RESONANT SPIN-FLAVOR PRECESSION SCENARIO}\\
\vspace{0.5cm}
{\large E.Kh. Akhmedov,
\footnote{On leave from Kurchatov Institute of Atomic
Energy, Moscow 123182, Russia}
\footnote{Also at Dipartimento di Fisica, Universit\`{a} di Roma "La Sapienza",
I.N.F.N.- Sezione di Roma, Italy}
\footnote{E-mail: akhmedov@tsmi19.sissa.it, ~akhm@jbivn.kiae.su}
A. Lanza, ~S.T. Petcov\footnote{Istituto Nazionale di Fisica Nucleare, Sezione
di Trieste, Italy}
\footnote{Permanent address: Institute of
Nuclear Research and Nuclear Energy, Bulgarian Academy of Sciences, BG-1784
Sofia, Bulgaria}}\\
\vspace{0.2cm}
{\em Scuola Internazionale Superiore di Studi Avanzati\\
Strada Costiera 11, I-34014 Trieste, Italy} \\
\end{center}
\vspace{0.4cm}
\begin{abstract}
We consider the implications of the recent results of SAGE and GALLEX
experiments for the solution of the solar neutrino problem in the framework
of the resonant neutrino spin-flavor precession scenario. It is shown that
this scenario is consistent with all the existing solar neutrino data
including the gallium results. The quality of the fit of the data depends
crucially on the magnetic field profile used which makes it possible to get
information about the magnetic field in the solar interior. In particular,
the magnetic field in the core of the sun must not be too strong ($\aprle 3
\times 10^6$ G). The detection rate in the gallium detectors turns out to be
especially sensitive to the magnitude of $\Delta m^2$. Predictions for
forthcoming solar-neutrino experiments are made.
\end{abstract}
\vspace{2cm}
\vspace{.5cm}
\end{titlepage}
\renewcommand{\thefootnote}{\arabic{footnote}}
\setcounter{footnote}{0}
\newpage
\section{Introduction}
One of the conceivable solutions of the solar neutrino problem
\cite{SNP1,SNP2,SNP3,GALLEX1} is related to the possible existence of magnetic
\cite{C,FSh,VV,VVO} or transition \cite{ShVa,VVO} magnetic moments of electron
neutrinos. If these magnetic moments are sufficiently large, a significant
fraction of the left-handed solar $\nu_e$ will precess into right-handed
neutrinos or antineutrinos  of the same or different flavor in the strong
toroidal magnetic
field of the convective zone of the sun. The resulting neutrinos are sterile
or almost sterile for currently operating neutrino detectors. It has been
pointed out \cite{AKHM1,LM,AKHM2} that the spin-flavor precession due to
transition magnetic moments of neutrinos, in which neutrino helicity and
flavor are rotated simultaneously, can be resonantly enhanced in matter, very
much in the same manner as the neutrino oscillations \cite{W,MS}. Neutrino spin
precession and resonant spin-flavor precession (RSFP) can explain the
observed deficiency of solar neutrinos \cite{SNP1,SNP2} as compared to the
predictions of the
standard solar model (SSM) \cite{BU,TCh,BP}. They can also account for time
variations of the solar
neutrino flux in anticorrelation with the solar activity for which there are
indications in the chlorine experiment of Davis and his collaborators
\cite{SNP1}. This follows from the drastic field-strength dependence of
the precession probability and the fact that the toroidal magnetic field of
the sun is strongest at maxima of solar activity.

Although the chlorine data seem to exhibit a strong time variation of
the $\nu_e$ detection rate $Q$, $Q_{max}/Q_{min}\approx 2-4$, no such time
dependence has been observed in the Kamiokande II experiment \cite{SNP2}.
The Kamiokande II results do not exclude, however, small ($\aprle 30\%$) time
variations of the solar neutrino signal.
The question of whether it is possible to reconcile large time variations
of the signal in the Homestake ${}^{37}\!$Cl experiment with small time
variations allowed by the water \v{C}erenkov detector has
been addressed in a number of papers \cite{AKHM7,AKHM5,BMR,OS,FY}. It has been
shown that the Homestake and Kamiokande data can be naturally reconciled
in the framework of the RSFP scenario. The key point is that the
Homestake and Kamiokande II detectors are sensitive to different regions of
the solar neutrino spectrum: the energy threshold in the Homestake experiment
is 0.814 MeV so that it is able to detect high energy ${}^{8}$B and the
intermediate energy ${}^{7}$Be and $pep$ neutrinos; at the same time the
energy threshold  in the Kamiokande II experiment was 7.5 MeV and so it was
only sensitive to the higher-energy fraction of the ${}^{8}$B neutrinos. Due to
the strong energy dependence of the probability of the RSFP,
the flux of lower-energy neutrinos can exhibit stronger time variations (i.e.
stronger magnetic field dependence) than the flux of the higher-energy ones.
Another point is that the Homestake experiment utilizes the
$\nu_{e}-{}^{37}\!$Cl charged current reaction, while in the Kamiokande
detector $\nu-e$ scattering which is mediated by both charged and
neutral-current weak interactions is used. If there are no sterile  neutrinos
the RSFP converts solar $\nu_{eL}$ into $\bar{\nu}_{\mu R}$ or
$\bar{\nu}_{\tau R}$ which are not observed in the chlorine detector but can
be detected (though with a smaller cross section than $\nu_e$)
in water \v{C}erenkov experiments. This also reduces the possible time
variation effect in the Kamiokande II detector.

In the present paper we perform a detailed numerical calculation of the
${}^{37}$Cl, Kamiokande and ${}^{71}$Ga detection rates in the framework of
the RSFP scenario and confront them with the existing solar neutrino data
including the recent results of the gallium experiments. We show that the
current solar neutrino observations can be perfectly well described on the
basis of the RSFP hypothesis.
The time dependence of the  detection rates turns out to be very sensitive
to the magnetic field profile used which can make it possible to discriminate
between various magnetic field configurations and get an information about
the magnetic field in the solar interior. We briefly also consider the
implications of our results for the forthcoming solar-neutrino experiments.
In particular, strong time variations of the ${}^7$Be neutrino signal is
predicted.

\section{General features of the RSFP}
For simplicity, we shall consider the RSFP of $\nu_e$ into $\bar{\nu}_{\mu}$
disregarding possible neutrino mixing and assuming that the direction of the
transverse magnetic field does not change along the neutrino trajectory. By
lifting  these assumptions one can enlarge the number of parameters of the
problem and therefore facilitate fitting the data; however, as we shall see,
a sufficiently good fit can be achieved without introducing additional
parameters.

We shall consider the Majorana neutrino case which is more promising for
the simultaneous fit of the Homestake and Kamiokande II data. Under the
assumptions made, $\nu_e$ and $\nu_{\mu}$ are neutrinos with definite
masses in vacuum ($m_1$ and $m_2$) and transition magnetic moment
$\mu_{e\mu}$. To make the paper self-contained, we recall here the main
features of the RSFP \cite{AKHM1,LM,AKHM2,AKHM3}. The evolution of the
$(\nu_{eL},~\bar{\nu}_{\mu R})$ system in matter and magnetic field is
described by the Schroedinger-like equation
\EQ
i\frac{d}{dt}\left(\begin{array}{c}\nu_{eL}\\ \bar{\nu}_{\mu R}\end{array}
\right)=\left(\begin{array}{cc}
\sqrt{2}G_F(N_e-N_n/2)+\frac{m_1^2}{2E} & \mu_{e\mu} B_{\bot}\\
\mu_{e\mu} B_{\bot} & \sqrt{2}G_F(N_n/2)+\frac{m_2^2}{2E}
\end{array}\right)
\left(\begin{array}{c}\nu_{eL}\\ \bar{\nu}_{\mu R}\end{array}\right)
\EN
Here $N_{e}$ and $N_{n}$ are the electron and neutron number densities and
$G_{F}$ is the Fermi constant. The mixing angle of $\nu_{eL}$ and
$\bar{\nu}_{\mu R}$ in matter and magnetic field is defined as
\EQ
\tan 2\theta_m = \frac{2\mu_{e\mu}B_{\bot}}{\sqrt{2}G_{F}(N_{e}-
N_{n})-\frac{\Delta m^2}{2E}}
\EN
where $\Delta m^2\equiv m_2^2-m_1^2$. The precession length $l$ is given by
\EQ
l=2\pi\left\{\left[\sqrt{2}G_{F}(N_{e}-N_{n})-\frac{\Delta m^2}
{2E}\right]^{2}+(2\mu_{e\mu}B_{\bot})^{2}\right\}^{-1/2}
\EN
The resonance density is a density at which the mixing angle $\theta_m$ becomes
$\pi/4$:
\EQ
\sqrt{2}G_{F}(N_{e}-N_{n})|_{res}=\frac{\Delta m^2}{2E}
\EN
The efficiency of the $\nu_{eL}\rightarrow \bar{\nu}_{\mu R}$ transition is
basically determined by the degree of its adiabaticity. The corresponding
adiabaticity parameter depends both on
the neutrino energy and the magnetic field strength at the resonance
$B_{\bot res}$:
\EQ
\lambda\equiv \pi \frac{\Delta r}{l_{res}}=
8\frac{E}{\Delta m^2}(\mu_{e\mu}B_{\bot res})^{2}L_{\rho}
\EN
Here $\Delta r$ is the resonance width, $l_{res}=\pi/\mu_{e\mu}B_{\bot res}$ is
the precession length at the resonance and $L_{\rho}\equiv|(N_e-N_n)_{res}/
\frac{d}{dt}(N_e-N_n)_{res}|$ is the characteristic
length over which matter density varies significantly in the sun. For the RSFP
transition to be almost complete, one should have $\lambda \gg1$; however, the
transitions can proceed with an appreciable probability even for $\lambda
\sim 1$.

In a non-uniform magnetic field, the field strength at resonance $B_{\bot res}$
depends on the resonance coordinate and so, through eq. (4), on the neutrino
energy. Thus, the energy dependence of the adiabaticity parameter
$\lambda$ in eq. (5) is, in general, more complicated than just $\lambda
\sim E$, and is determined also by the magnetic field profile inside the sun.
\section{Calculation of detection rates}
We have numerically integrated eq. (1) to find the $\nu_{eL}$
survival probability in the sun. It was then used to calculate the neutrino
detection rates for the
Homestake, Kamiokande II and gallium experiments utilizing the solar density
profiles, solar neutrino spectra and cross sections of $\nu_e$ capture on
${}^{37}\!$Cl and ${}^{71}$Ga from ref. \cite{BU}, as well as realistic
detection efficiency of Kamiokande II experiment and standard-model $\nu-e$
cross sections.

Unfortunately, very little is known about the structure of the magnetic field
inside the sun so that one is forced to use various more or less "plausible"
magnetic field configurations. In our calculations we assumed the transverse
magnetic field profiles to consist of two separate spherically symmetric
pieces: constant nonuniform
internal magnetic field with its scale defined by the
parameter $B_1$ and magnetic field in the convective zone $B_{c}(x)$,
characterized by the overall scale $B_0$ which can vary in time with changing
solar activity \cite{AkBy,AKHM5}:
\EQ
B_{\bot}(x)=\left\{\begin{array}{c}
B_1\left(\frac{0.1}{0.1+x}\right)^2,~~~0\leq x<x_0,\\
B_c(x),~~~x_0\leq x<x_{max}
\end{array}\right.
\EN
where $x\equiv r/R_{\odot}$, $r$ being the distance from the center of the
sun and $R_{\odot}$ being the solar radius. For the convective-zone magnetic
field $B_c(x)$ we used the following profiles:
\EQ
B_{c}(x)=\left\{\begin{array}{c}
B_0\frac{x-x_{0}}{x_c-x_{0}},~~~x_{0}\leq x<x_c,\\
B_0-(B_0-B_f)\frac{x-x_{c}}{1-x_{c}},~~~x_{c}\leq x\leq 1
\end{array}\right.
\EN
with $x_{0}=0.65$, 0.7 and 0.75, $x_c=0.75$ and 0.85, $B_f=0$ and 100 G,
\EQ
B_{c}^{(n)}(x)=B_0\left[1-\left(\frac{x-0.7}{0.3}\right)^n\right],
{}~~~x_{max}=1
\EN
with $n$=2, 6 and 8, and
\EQ
B_{c}(x)=B_0/\{1+\exp[(x-0.95)/0.01]\},~~~x_{max}=1.3.
\EN

It is not clear whether or not there is a strong magnetic field in the inner
regions of the sun and so one has no idea about the possible magnitude of
$B_1$; in our calculations we treated it as a free parameter. As to the
maximum value of the convective zone magnetic field, fields of the order of a
few kG or a few tens of kG are usually considered possible. There are,
however, arguments based on the helioseismology data that it can achieve
as large values as a few MG \cite{D}. The typical values of $B_0$ we used were
of the order of a few tens of kG.

The magnetic field strength enters the evolution equation (1) being
multiplied by the neutrino transition magnetic moment $\mu_{e\mu}$.
The existing upper limits on the magnetic moment of the electron neutrino
include the laboratory bound $\mu_{\nu_e}\le (3-4)\times 10^{-10}\mu_{B}$
($\mu_B$ is the electron Bohr magneton) from the reactor $\bar{\nu}_e$
experiments \cite{react}, as well as astrophysical, cosmological and
supernova 1987A bounds. The cosmological and supernova bounds do not apply
to transition magnetic moments of Majorana neutrinos which are of major
interest to us. The astrophysical bounds come from the limits on the energy
loss rates of white dwarfs and helium stars, which yield $\mu_{\nu_e}\le
10^{-11}\mu_{B}$ \cite{WD}; recently even more stringent bounds have been
obtained from the analysis of the data on helium stars, $\mu_{\nu_e}\le 3
\times 10^{-12}\mu_{B}$ \cite{Raffelt} and $\mu_{\nu_e}\le 10^{-12}\mu_{B}$
\cite{CDI}. One should notice that besides the upper limits on neutrino
magnetic moments there is also an astrophysical "prediction" for the transition
magnetic moment $\mu_{\nu_e\nu_{\tau}}$ (or $\mu_{\nu_e\nu_{\mu}})
\simeq 10^{-14}\mu_{B}$ from the decaying neutrino hypothesis
\cite{Dennis}. For neutrino magnetic moments of this magnitude to be relevant
to the solar neutrino problem, one would need convective zone magnetic field of
the order of 10 MG. In our calculations we have taken $\mu_{e\mu}=10^{-11}
\mu_B$ where $\mu_B$ is the electron Bohr magneton. The results for any other
value of $\mu_{e\mu}$ can be readily obtained by simply rescaling the magnetic
field strength.
\section{Results and discussion}
Let us first recall the experimental situation. The observed chlorine
detection rates in the minima of solar activity were $4.1\pm 0.9$ SNU (cycle
21) and $4.2\pm 0.75$ SNU (cycle 22)
\footnote{1 SNU (Solar Neutrino
Unit)=$10^{-36}$ events/target atom/s.}; those in solar maxima were $0.4\pm
0.2$ SNU (cycle 21) and $1.2\pm 0.6$ SNU (cycle 22). The average detection
rate in the chlorine experiment is $Q_{ave}=2.28\pm 0.23$ SNU \cite{SNP1}.
This should be compared with the SSM prediction $(Q_{{\rm SSM}})_{{\rm Cl}}=
8.0\pm 3.0$ SNU (effective "3$\sigma$" error) \cite{BP}.

The Kamiokande II Collaboration reported the value for the ratio of the
observed and the SSM
detection rates $R\equiv Q/Q_{\rm SSM}=0.46\pm 0.05$(stat.)$\pm 0.06$(syst.),
the allowed time variation of $R$b being $\aprle 30\%$ \cite{SNP2}. Taking into
account the first reported data of the Kamiokande III experiment would just
shift the central value of $R$ to 0.49 \cite{KamIII}. As was emphasized
above, there was no indication of possible time dependence of the signal
in the Kamiokande II experiment; however, time variation in the range
$R\simeq 0.3-0.7$ is not excluded (at the $1\sigma$ level).

By now, the results of two gallium solar neutrino experiments are available.
The combined SAGE result for 1990-1991 is $58^{+17}_{-24}$(stat.)$\pm
14$(syst.) SNU which is in agreement with the GALLEX data for
1991-1992 of $83\pm 19$(stat.)$\pm 8$(syst.) SNU \cite{GALLEX1}.

It is interesting to note that the SAGE results for 1990 and 1991 were
$20^{+15}_{-20}\pm 32$ and $85^{+22}_{-32}\pm 20$ SNU respectively, i.e.
they exhibited the tendency to increase with time while the solar activity
was decreasing. Although this is by no means sufficient to conclude that
the observed signal was varying in time, the data compare favorably with
such an assumption.

Let us now turn to the results of our calculations which are presented in figs.
1--5. It can be seen from fig. 1 that for the "triangle"
convective-zone magnetic field of eq. (7) with $x_0$=0.70, $x_c=0.85$,
$x_{max}=1$, $B_f=0$ and no inner field, all the available solar-neutrino data
can be perfectly well fitted for $\Delta m^2 \sim 7\times 10^{-9}$ eV${}^2$.
For the parameter of the convective-zone magnetic field $B_0$ varying between
25 and 50 kG, the detection rate in the chlorine experiment changes between
4.5 and 1.5 SNU. For the same range of $B_0$ the ratio $R$ measured in the
Kamiokande II experiment varies between 0.65 and 0.35, which
is within the experimentally allowed domain. The calculated germanium
production rate in gallium experiments varies between 100 and 73 SNU; the
latter value, corresponding to high solar activity periods, is in a very good
agreement with both SAGE and GALLEX results.

The combined results of the calculations of the ${}^{37}\!$Cl, Kamiokande
and ${}^{71}$Ga detection rates are presented in the form of iso-SNU
(iso-suppression) contour plot in fig. 2. In this figure, the shaded areas
correspond
to the allowed values of the parameters which are chosen so that the maximum
and minimum detection rates in the chlorine experiment range, respectively,
between 3.6 and 4.8 SNU, and 1.9 and 1.5 SNU. For the gallium signal at the
maximum of solar activity, the whole range between 62 and 104 SNU was allowed.
This corresponds to $1\sigma$ errors except for the minimum signal in the
chlorine data $(Q_{\rm Cl})_{min}$, for which the quoted errors seem to be
underestimated \cite{Krauss}, and therefore we considered as large
$(Q_{\rm Cl})_{min}$ as 1.9 SNU possible; we were unable to get values of
$(Q_{\rm Cl})$ below 1.5 SNU with the magnetic field configurations we
used\footnote{This, of course, is only true if we want to fit all three data
sets simultaneously.}. Notice that the available gallium results should be
considered as "low-signal" ones since both SAGE and GALLEX were in fact taking
data during the period of high solar activity. For the Kamiokande experiment,
it is difficult to choose the allowed ranges of $R$ for high and low solar
activity periods. We have taken the maximum signal/SSM ratio to vary in the
range 0.58--0.68, and minimum one, in the range 0.3--0.4. The numbers 0.68
and 0.30 correspond to the observed maximal and minimal values of $R$
(including 1$\sigma$ errors for the corresponding runs); the width of the
allowed $R$ regions $\Delta R=0.1$ was taken rather arbitrarily.

The allowed ranges of the parameters which follow from fig. 2
correspond to
\footnote{The shaded areas at larger values of $B_0$ correspond to
direct correlation of neutrino signals with solar activity instead of
anticorrelation.}.
\EQ
\Delta m^2 \simeq (4\times 10^{-9}-2 \times 10^{-8})~
{\rm eV^2},~~ (B_0)_{min}\simeq 28~{\rm kG},~~ (B_0)_{max}\simeq 50~{\rm kG}
\EN
We see from the figure that for the high signal (low
$B_0$) domain, the ${}^{37}\!$Cl and Kamiokande data mainly constrain the
allowed magnitudes of the convective-zone magnetic field $B_0$, i.e.
the horizontal size of the allowed strip. At the same time, for certain
values of $B_0$ the ${}^{71}$Ga data constrain (from below) the vertical size
of the strip, i.e. the allowed values of $\Delta m^2$.
For the low signal ($B_0 \sim 50$ kG) domain of the allowed parameter space,
the
values of $B_0$ are constrained by the chlorine data while the upper and the
lower limits on $\Delta m^2$ are determined by the Kamiokande and gallium data
respectively. This shows that the gallium experiments are
especially sensitive to the magnitude of $\Delta m^2$.

Similar results are obtained if one takes $x_0=0.75$ instead of 0.70
in the magnetic field profile of eq. (7). All the other magnetic-field
configurations
that we used in our calculations either resulted in very poor fits of the
experimental results or completely failed to reproduce the chlorine, Kamiokande
and gallium data simultaneously.
For example, one can fit the data with the magnetic field
in the convective zone of eq. (8) with $n=6$ provided there is a moderate
inner field ($B_{1}=2\times 10^6$ G). However, the allowed range of
$\Delta m^2$ turns out to be extremely narrow (fig. 3) and is likely to
disappear with improving accuracy of the data.

In refs. \cite{BMR} and \cite{OS} numerical calculations of the Homestake and
Kamiokande detection rates in the framework of the RSFP scenario have been
performed. Satisfactory description of the results of these experiments has
been achieved, but the predicted gallium signal in solar maximum for the
parameter range they found was $\sim 20-30$ SNU which is considerably lower
than the positive results obtained by SAGE and GALLEX Collaborations.
It is easy to understand why the gallium detection rates in the periods of
high solar activity predicted in refs. \cite{BMR,OS} were too small, whereas
the results of our calculations are in good agreement with the data. The main
contribution to the signal in the gallium experiments comes from the
low-energy $pp$ neutrinos which, according to eq. (4), encounter the resonance
at higher density than the ${}^7$Be and ${}^8$B neutrinos. For them the
resonant region can lie in the radiation zone and the efficiency of the RSFP
will drastically depend on the magnitude of the inner magnetic field of the
sun. If this field is absent, as in the calculations the results of which are
shown in figs. 1 and 2, or just not too
strong, as in the field configuration of eq. (8) with $n$=6 and $B_1=2\times
10^{6}$ G, the $pp$ neutrinos either leave
the sun unscathed or experience very weak conversion. Therefore, in this case
one can expect unsuppressed $pp$ signal in the gallium experiments in good
agreement with observations. Note that the authors of ref. \cite{BMR} assumed
very strong inner magnetic field ($B_1=2\times 10^7$ G).

The resonant density depends on the magnitude of $\Delta m^2$ as well, so
for small enough $\Delta m^2$ the $pp$ neutrinos can also undergo resonant
transitions in the convective zone. As a consequence, the $pp$ neutrino flux
can be heavily suppressed even in the absence of strong inner magnetic field.
In this case the ${}^7$Be, $pep$
and ${}^8$B neutrinos will undergo resonant conversion closer to the
surface of the sun and the Homestake and Kamiokande II data can only be
reproduced if the convective-zone magnetic field $B_{c}(r)$ does not
decrease too fast with increasing $r$. In ref. \cite{OS} the value of
$\Delta m^2$ was chosen to be rather small ($\Delta m^2 \simeq 5 \times
10^{-9}$ eV${}^2$) so that a significant fraction of $pp$ neutrinos
experienced resonant transition in the convective zone.

{}From the above results it follows that

(1) in the vicinity of $\Delta m^2 \simeq (6-8)\times 10^{-8}$ eV${}^2$
the gallium detection rate should exhibit strong dependence on the value of
$\Delta m^2$ since for smaller $\Delta m^2$ an appreciable fraction of $pp$
neutrinos will have the resonance in the convective zone, whereas for larger
$\Delta m^2$ most of them will encounter the resonance in the radiation zone
where the magnetic field need not be strong\footnote{The border-line value
of $\Delta m^2$ depends, of course, on whether or not the magnetic field
is strong near the bottom of the convective zone. For our field
configurations of eq. (7) for which $B_{\bot}(x)$ achieves its maximum in
the center of the convective zone this value can be somewhat smaller.};

(2) the gallium detection rate is very sensitive to the magnitude of the
inner magnetic field of the sun and can be used to set an upper limit
on it;

(3) while Homestake and Kamiokande II data can well be reproduced with
a wide class of magnetic field configurations, the simultaneous fit of the
results of all three kinds of experiments, the chlorine, water \v{C}erenkov
and the gallium ones, is very sensitive to the solar magnetic field profile
used.  This stems from the fact that these experiments are sensitive to
neutrinos of different energies which experience the RSFP transitions at
different distances from the center of the sun and therefore probe the
magnetic field distribution $B_{\bot}(r)$ in the wide range of $r$.

The first point is illustrated by figs. 1c and 2 which show that the
dependence of $Q_{\rm Ga}$ on $\Delta m^2$ is much stronger than in the case
of chlorine
and Kamiokande detection rates. The second point is illustrated by fig. 4
in which all the three detection rates are shown as functions of $B_{1}$
and $B_{0}$ for a fixed value of $\Delta m^2$. One can see that for $B_{1}
\aprge 3\times 10^6$ G the predicted gallium detection rate is too small. The
third point is confirmed by our calculations with various magnetic field
profiles. As we already mentioned, only two magnetic field configurations gave
acceptable fits of the data. With most other configurations it was not
difficult to fit the chlorine and Kamiokande results, but not all three
data sets simultaneously.

In fig. 5 the magnetic field dependence of the suppression factors
of the $pp$, $pep$, ${}^7$Be and ${}^8$B neutrino contributions to the
gallium detection rate is illustrated (the suppression factors for the $pep$,
${}^7$Be and ${}^8$B neutrino contributions to $Q_{\rm Cl}$ are practically
the same). One can see that the ${}^7$Be flux is expected to exhibit a strong
time variation, between $\simeq 0.6 \times$(SSM prediction) in the low solar
activity periods and $\simeq 0.1 \times$(SSM prediction) at high activity.
This should be
clearly seen in the forthcoming Borexino experiment which is intended for
detecting the ${}^7$Be neutrino signal. Strong time dependence of the ${}^7$Be
neutrino flux is what one expects on general grounds from the condition
of having stronger time variations for the chlorine detection rate than for
the Kamiokande one\footnote{One should note that for our magnetic field
configuration of eq. (8) with $n=6$ and $B_1=2\times 10^{6}$ G the predicted
${}^7$Be flux is almost constant since the $\Delta m^2$ value which fits the
data is rather large and the ${}^7$Be neutrinos encounter the resonance in
the radiation zone where the magnetic field is constant in time. However, as
we already mentioned, the allowed $\Delta m^2$ range is extremely narrow in
this case.}.

Let us briefly discuss the predictions for the other solar neutrino
experiments. In the gallium detectors, the signal should experience
moderate time variation, the signal at minimum solar activity being
about 100-110 SNU. In the SNO experiment, time variation of the signal in
the charged-current channel should be similar to that in the chlorine
detector; in addition, the neutrino spectrum measured in this channel should
be distorted. In the neutral-current channel one should see unsuppressed
{\em constant} signal (the signal can be somewhat suppressed if the
SSM overestimates the ${}^8$B flux or if there are neutrino oscillations
along with the RSFP).

It is interesting to note that in order to reproduce the existing
solar neutrino data, one
needs a modest variation of the magnetic field parameter $B_0$, only
by a factor of two or so. In reality, the situation can be much more
complicated. It is possible that the maximum of the convective-zone
magnetic field does not just change in time having fixed spatial
position, but rather floats from
the bottom of the convective zone to its surface during the solar cycle
\cite{ZR}. We plan to consider this effect as well as more realistic
3-dimensional magnetic field configurations in a subsequent publication.

As we emphasized above, only the Homestake data seem to exhibit a time
variation of the neutrino signal whereas the Kamiokande data do not support
this possibility. In our calculations we just fitted the upper
and lower
values of $Q_{\rm Cl}$ having in mind that the parameter sets which fit
these values for some $(B_{0})_{min}$ and $(B_{0})_{max}$ will also fit all
the intermediate $Q_{\rm Cl}$ for the magnitudes of $B_{0}$ lying in between
these values. A more consistent approach is to perform a $\chi^2$ fit of all
the data sets on the run-by-run basis, as, for example, it has been done for
Homestake and Kamiokande II data in \cite{Bahcall,Krauss}. The results of
such an analysis will be presented elsewhere.

In conclusion, we have shown that the results of the ${}^{37}\!$Cl, Kamiokande
II and ${}^{71}$Ga solar neutrino experiments can be reproduced assuming that
the solar $\nu_e$ undergo resonant spin-flavor precession into
$\bar{\nu}_{\mu}$ or $\bar{\nu}_{\tau}$ in the magnetic field of the sun.
The gallium detection rate turns out to be especially sensitive to
the magnitude of the parameter $\Delta m^2$ . It also depends drastically on
the strength of the inner magnetic field of the sun. Using the positive results
of SAGE and GALLEX experiments, we have obtained an upper limit on the
strength of such a field: $B_{1}\aprle 3 \times 10^6$ G. Our calculations
show that the quality of the
combined fit of the data is very sensitive to the magnetic field configuration
used, which opens up the possibility of extracting information on the
strength and profile of the solar magnetic field from solar neutrino data.
We predict very strong time variation of the ${}^7$Be neutrino flux which
should be observable in the Borexino experiment. \\

We would like to thank D. Sciama for his encouragement and support. The
work of S.T.P. was supported in part by the Bulgarian National Science
Foundation via grant PH-16.\\

\noindent {\em Note added}. When this paper was being typed we received a
preprint by Krastev
\cite{Plamen} in which analogous analysis has been done assuming that the
direction of the solar magnetic field varies along the neutrino trajectory.
It was shown that in this case a satisfactory description of the data can
also be achieved.

\newpage
\centerline{\bf \large Figure captions}

\vskip 1truecm
\noindent
Fig. 1. Predicted detection rates for the chlorine (a), Kamiokande (b) and
gallium (c) experiments as functions of convective zone magnetic field
parameter $B_0$ for the "triangle" magnetic field configuration of eq. (7)
with $x_0=0.70$, $x_c=0.85$, $x_{max}=1$, $B_f=0$ and no inner magnetic field;
$\mu_{e\mu}=10^{-11}\mu_B$ is assumed. The numbers near the curves indicate
the values of $\Delta m^2$ in units of eV${}^2$. The groups of closely
located curves in figs. 1a and 1b correspond to the values of $\Delta
m^2$ in the range $5\times 10^{-9}-8\times 10^{-9}~{\rm eV^2}$.

\vskip 1truecm
\noindent
Fig. 2. The iso-SNU (iso-suppression) contours for ${}^{37}\!$Cl, Kamiokande
and ${}^{71}$Ga experiments in the $(B_0,~\Delta m^2)$ plane. The magnetic
field
configuration is the same as in fig. 1. The full lines are chlorine iso-SNU
curves (1.5, 1.9, 2.2, 3.6, 4.8, 5.2 and 5.8 SNU), the dotted lines
correspond to the ratio $R$=0.30, 0.40, 0.58 and 0.68 for the Kamiokande
experiment, and the dash-dotted lines represent the gallium iso-SNU curves
(62.0, 83.0, 104.0 and 120 SNU). The shaded areas show the allowed ranges of
parameters (see the text).

\vskip 1truecm
\noindent
Fig. 3. Same as fig. 2, but for the magnetic field configuration of eq. (8)
with $n=6$ and $B_1=2\times 10^6$ G.

\vskip 1truecm
\noindent
Fig. 4. The iso-SNU (iso-suppression) contours in the ($B_0,~B_1$) plane.
The magnetic field configuration and the definition of the curves are the same
as
in fig. 2.

\vskip 1truecm
\noindent
Fig. 5. Suppression factors for neutrinos from the different sources for the
gallium experiments as functions of the convective zone magnetic field
parameter $B_0$. The magnetic field configuration is the same as that
used for fig. 1.
\end{document}